\documentclass[aps,prd,onecolumn,groupedaddress,showpacs,nofootinbib,amssymb]{revtex4}
\usepackage[dvips]{graphicx,color}
\usepackage{amssymb}
\usepackage{amsmath}
\usepackage{graphicx}
\usepackage{amsfonts}

\newcommand{\be}{\begin{equation}}
\newcommand{\ee}{\end{equation}}
\newcommand{\bea}{\begin{eqnarray}}
\newcommand{\eea}{\end{eqnarray}}
\newcommand{\beaa}{\begin{eqnarray*}}
\newcommand{\eeaa}{\end{eqnarray*}}

\begin{document}

\title{Loop Quantum Cosmology Scalar Field Models}
\author{
K. Kleidis,$^{1}$\,\thanks{kleidis@teiser.gr}
V.K. Oikonomou$^{1}$,\,\thanks{v.k.oikonomou1979@gmail.com}}
\affiliation{
$^{1)}$ Department of Mechanical Engineering\\ Technological
Education Institute of Central Macedonia \\
62124 Serres, Greece
}

\begin{abstract}
In this work we use the Loop Quantum Cosmology modified scalar-tensor reconstruction techniques in order to investigate how bouncing and inflationary cosmologies can be realized. With regard to the inflationary cosmologies, we shall be interested in realizing the intermediate inflation and the Type IV singular inflation, while with regard to bouncing cosmologies, we shall realize the superbounce and the symmetric bounce. In all the cases, we shall find the kinetic term of the LQC holonomy corrected scalar-tensor theory and the corresponding scalar potential. In addition, we shall include a study of the effective equation of state, emphasizing at the early and late time eras. As we demonstrate, in some cases it is possible to have a nearly de Sitter equation of state at the late-time era, a result that could be interpreted as the description of a late-time acceleration era. Also, in all cases we shall examine the dynamical stability of the LQC holonomy corrected scalar-tensor theory, and we shall confront the results with those coming from the corresponding classical dynamical stability theory. The most appealing cosmological scenario is that of a Type IV singular inflationary scenario, in which the singularity may occur at the late-time era. As we demonstrate, for this model, during the dark energy era, a transition from non-phantom to a phantom dark energy era occurs.
\end{abstract}

\pacs{04.50.Kd, 95.36.+x, 98.80.-k, 98.80.Cq,11.25.-w}

\maketitle

\section{Introduction}

Admittedly, the late-time acceleration of our Universe, firstly observed in the late 90's \cite{riess}, is one of the striking surprises that the Universe offered to the scientific community, if not the most striking one. This late-time acceleration era along with the primordial era and the dark matter problem, constitute three of the main puzzles in modern cosmology. Indeed, the ultimate goal is to find a consistent way to model this late-time acceleration, which is attributed to the so-called dark energy. Also, the primordial era can also be described by inflationary theories \cite{inflation1,inflation2,inflation3,inflation4,inflation5,inflation6}, and by bouncing cosmologies \cite{brande1,bounce1,bounce1a,bounce1b,bounce2,bounce3,bounce4,bounce5,matterbounce1,matterbounce2,matterbounce3,matterbounce4,matterbounce5,matterbounce6,matterbounce7,Nojiri:2016ygo,Odintsov:2015ynk,Oikonomou:2015qha,Makarenko:2017vuk,Odintsov:2015zza}, which both can produced a blue-tilted spectral index of primordial curvature perturbations, which is recently has be verified by the Planck observational data \cite{planck}. With regards to the dark matter problem, it is a longstanding problem in theoretical physics, and there are various particle \cite{Oikonomou:2006mh} or geometrical \cite{Chamseddine:2013kea,Nojiri:2014zqa} proposals that may address this problem in a self-consistent way, however no evidence of a particle dark matter exists up to date. With regard to the early and late-time acceleration eras, it is possible to describe these in a unified way, in the context of modified gravity \cite{inflation1,reviews2,reviews3,reviews4}, and this kind of description has firstly appeared in Ref. \cite{Nojiri:2003ft}. It is noteworthy to mention that perfect or viscous fluid approaches \cite{Brevik:2017msy,Elizalde:2017dmu,Elizalde:2016tmn,Brevik:2017juz}, can also describe the early and late-time acceleration eras in a unified way.

Apart from the modified gravity or fluid descriptions of late and early-time acceleration, the scalar field description is also frequently used, and in many cases, canonical or non-canonical scalar fields are used \cite{scalarrecon0,scalarrecon1,scalarrecon2,scalarrecon3,scalarrecon4,scalarrecon5,scalarrecon6,scalarrecon7,scalarrecon8,scalarrecon9}, which in some cases can turn into phantom fields. For scalar fields, there exist several reconstruction techniques, which can be used to realize various cosmological evolutions \cite{scalarrecon0,scalarrecon1,scalarrecon2,scalarrecon3,scalarrecon4,scalarrecon5,scalarrecon6,scalarrecon7,scalarrecon8,scalarrecon9}. In a recent work \cite{Oikonomou:2016phk}, the scalar-tensor reconstruction techniques were extended in the context of Loop Quantum Cosmology (LQC) \cite{LQC1,LQC2,LQC3,LQC4,LQC5,LQC6,LQC7,LQC8,LQC9,LQC10,LQC11,LQC12,LQC13,LQC14,LQC15,LQC16,LQC17,LQC18,LQC19,Odintsov:2016apy,Oikonomou:2017mlk,Kleidis:2017ftt}, and the first effects of holonomy corrected LQC were investigated. One of the striking results was that no Rip singularities occur if the holonomy corrections are  taken into account in the theory. As was demonstrated in \cite{Oikonomou:2016phk}, the LQC-corrected scalar-tensor theory coincides with the classical results \cite{scalarrecon0,scalarrecon1,scalarrecon2,scalarrecon3,scalarrecon4,scalarrecon5,scalarrecon6}, in the classical limit. In fact, the quantum effects are quantified in terms of the a parameter called critical density, denoted as $\rho_c$, and the classical limit is achieved when this parameter tends to infinity, that is $\rho_c\to \infty$. In some sense, this is the most important feature of LQC, since one can have a solid idea of the first quantum effects on the classical theory, which to our opinion is invaluable since the full quantum gravity theory is lacking.

In this paper we shall investigate how various cosmologies can be realized by using LQC-corrected scalar-tensor theories. By using the reconstruction scheme of \cite{Oikonomou:2016phk}, we shall realize four quite popular cosmologies, namely the intermediate inflation scenario \cite{Barrow:1990td,Barrow:1993zq,Barrow:2006dh,Barrow:2014fsa}, the Type IV singular inflation scenario \cite{Odintsov:2015gba,Nojiri:2015fra,Nojiri:2015fia,Oikonomou:2015qfh}, the symmetric bounce scenario, and the superbounce scenario \cite{Odintsov:2015uca}. In all cases we shall find the exact analytic form of the LQC-corrected kinetic term and scalar potential. In addition, we shall investigate the behavior of the LQC-corrected effective equation of state, emphasizing at the late and early-time eras, for the inflationary scenarios, and with regard to the bounces, we emphasize to the era near the bouncing point. As we demonstrate, it is possible to consistently describe a late-time acceleration era. Finally, we examine the dynamical stability of each model, again emphasizing at cosmic times of interest, and we compare the results with the classical dynamical stability results.

This paper is organized as follows: In section II, we present in brief the essential features of LQC-corrected scalar-tensor theory, and in section III we use this formalism to find the LQC-corrected scalar-tensor description of the theories which realize the intermediate and Type IV singular inflationary scenarios, and also the superbounce and symmetric bounce scenarios. In section IV we study the dynamical stability of the LQC-corrected dynamical cosmological equations, and we also compare the results with the classical picture. Finally, the conclusions follow in the end of the paper.

Before we start off our analysis, we need to note that in this study we shall assume that the geometric background is a flat Friedmann-Robertson-Walker (FRW) geometric background, which has the line element,
\be
\label{metricfrw} ds^2 = - dt^2 + a(t)^2 \sum_{i=1,2,3}
\left(dx^i\right)^2\, ,
\ee
where $a(t)$ is the scale factor. Also the Ricci scalar in this geometry is equal to,
\begin{equation}\label{ricciscalar}
R=12H^2+6\dot{H}\, ,
\end{equation}
where $H=\frac{\dot{a}}{a}$, is the Hubble rate.

\section{Essential Features of LQC-corrected Scalar-Tensor Reconstruction Method}

In this section we briefly review the formalism of LQC-corrected scalar-tensor gravity developed in Ref. \cite{Oikonomou:2016phk}, where details on the material we shall present can be found. In the context of LQC \cite{LQC1,LQC2,LQC3,LQC4,LQC5,LQC6,LQC7,LQC8,LQC9,LQC10,LQC11,LQC12,LQC13,LQC14,LQC15,LQC16,LQC17,LQC18,LQC19}, the Hamiltonian that captures the quantum effects in the Universe is,
\begin{equation}\label{effhamilt}
\mathcal{H}_{LQC}=-3V\frac{\sin^2(\lambda \beta)}{\gamma^2\lambda^2}+V\rho\, ,
\end{equation}
with $\gamma$ being the Barbero-Immirzi parameter, $\lambda$ being a parameter with dimensions of length, $V$ being the  world volume $V=a(t)^3$, where $a(t)$ is the scale factor of the Universe, and finally $\rho$  being the total energy density of the matter fluids present in the Universe. By taking into account the Hamiltonian constraint $\mathcal{H}_{LQC}=0$, we obtain,
\begin{equation}\label{hamiltonianconstr}
\frac{\sin^2(\lambda \beta)}{\gamma^2\lambda^2}=\frac{\rho}{3}\, ,
\end{equation}
and also by using the anti-commutation identity,
\begin{equation}\label{anticom}
\dot{V}=\{V,\mathcal{H}_{LQC}\}=-\frac{\gamma}{2}\frac{\partial \mathcal{H}_{LQC}}{\partial \beta},
\end{equation}
we obtain the holonomy LQC-corrected Friedmann equation \cite{LQC1,LQC2,LQC3,LQC4,LQC5,LQC6,LQC7,LQC9,LQC10,LQC11,LQC12,LQC13,LQC14,LQC15,LQC16,LQC17,LQC18,LQC19},
\begin{equation}\label{holcor1}
H^2=\frac{\kappa^2\rho}{3}\left (1-\frac{\rho}{\rho_c}\right )\, .
\end{equation}
The LQC-corrected Friedmann equation (\ref{holcor1}) is one of the most appealing attributes of LQC, since it entails all the quantum aspects of the theory, and also the classical theory can be obtained from it. Indeed, the quantum effects are due to the second term in the right-hand side of Eq. (\ref{holcor1}), which depends on the parameter $\rho_c$, which is called critical density. The classical limit in the theory can be obtained if the limit $\rho_c\to \infty$ is taken, so for finite $\rho_c$, the quantum effects of LQC on the classical theory are taken into account.

The total matter fluid energy density $\rho (t)$ satisfies the energy conservation identity,
\begin{equation}\label{cont}
\dot{\rho}(t)=-3H\Big{(}\rho(t)+P(t) \Big{)}\, ,
\end{equation}
with $P(t)$ being the total effective pressure of the matter fields. By combining Eqs. (\ref{cont}) and (\ref{holcor1}), after some algebra we obtain,
\begin{equation}\label{eqnm}
\dot{H}=-\frac{\kappa^2}{2}(\rho+P)(1-2\frac{\rho}{\rho_c})\, .
\end{equation}
Notice now that if the classical limit is taken, the equations (\ref{holcor1}) and (\ref{eqnm}) are modified as follows,
\begin{equation}\label{classicaleqns}
\dot{H}=-\frac{\kappa^2}{2}(\rho+P),\,\,\,H^2=\frac{\kappa^2\rho}{3}\, ,
\end{equation}
and notice that these are actually the classical equations. Let us proceed to see how the reconstruction formalism works in the LQC case, so by solving Eq. (\ref{holcor1}), with respect to $\rho$, we obtain,
\begin{equation}\label{sol1}
\rho=\frac{\kappa ^2 \rho_c\pm \sqrt{-12 H^2 \kappa ^2 \rho_c+\kappa ^4 \rho_c^2}}{2 \kappa ^2}\, ,
\end{equation}
and at this point one needs to decide which sign to keep in the above equation. This will be revealed by the condition that the above equation must yield the correct classical limit. The total pressure is,
\begin{equation}\label{pr}
P(t)=-\rho-\frac{2\dot{H}}{\kappa^2(1-\frac{2\rho}{\rho_c})}\, ,
\end{equation}
in effect, the Equation of State (EoS) parameter $w_{eff}$ is equal to,
\begin{equation}\label{eosdef}
w_{eff}=-1-\frac{\dot{H}}{3H^2}\pm \frac{\rho_c \dot{H}}{3 H^2 \sqrt{\rho_c \left(\rho_c-12 H^2\right)}}\, ,
\end{equation}
with the plus sign corresponding to the plus sign in Eq. (\ref{sol1}) and in addition, the minus in (\ref{eosdef}) corresponds to the minus in Eq. (\ref{sol1}). If the classical limit is taken, that is, when $\rho_c\to \infty$, the EoS $w_{eff}$ must coincide with the classical result, which is,
\begin{equation}\label{claseos}
w_{eff}=-1-\frac{2\dot{H}}{3H^2}\, ,
\end{equation}
so only the minus sign has to be kept in Eq. (\ref{sol1}), so we finally have,

Obviously, when , the EoS parameter must be equal to the classical result \cite{scalarrecon0,scalarrecon1,scalarrecon2},
\begin{equation}\label{rhodefcorrect}
\rho=\frac{\kappa ^2 \rho_c- \sqrt{-12 H^2 \kappa ^2 \rho_c+\kappa ^4 \rho_c^2}}{2 \kappa ^2}\, ,
\end{equation}
and in effect, the LQC-corrected EoS is,
\begin{equation}\label{eosdefcorrect}
w_{eff}=-1-\frac{\dot{H}}{3H^2}- \frac{\rho_c \dot{H}}{3 H^2 \sqrt{\rho_c \left(\rho_c-12 H^2\right)}}\, .
\end{equation}
Another problem comes up if the limit $\rho_c\to \infty$ is taken in Eq. (\ref{rhodefcorrect}), since we obtain $\rho=0$, which is not the classical result. The problem arises from the fact that the limit $\rho_c\to 0$ cannot be taken directly in Eq. (\ref{rhodefcorrect}), so by combining Eqs. (\ref{holcor1}) and (\ref{rhodefcorrect}), we rewrite $\rho$ in the following way,
\begin{equation}\label{correctdef}
\rho=\frac{3H^2}{\kappa^2\left( 1-\frac{\kappa ^2 \rho_c- \sqrt{-12 H^2 \kappa ^2 \rho_c+\kappa ^4 \rho_c^2}}{2 \kappa ^2\rho_c}\right)}\, .
\end{equation}
Now the above equation yields the correct result in the limit $\rho_c\to \infty$, which is $H^2=\frac{\kappa^2\rho}{3}$. Before we proceed, let us highlight the basic equations which will constitute our reconstruction technique, so hereafter the energy density will be that of Eq. (\ref{correctdef}), while the total pressure will be that of Eq. (\ref{pr}), and finally, the EoS parameter will be that of Eq. (\ref{eosdef}).

In the following we shall assume a FRW metric of the form (\ref{metricfrw}), and we shall assume that the matter fluids consists only from a non-canonical scalar field, the energy density $\rho_{\phi}$ and pressure $P_{\phi}$ of which, are,
\begin{equation}\label{presenrg}
\rho_{\phi}=\frac{1}{2}\omega(\phi)\dot{\phi}^2+V(\phi),\,\,\,P_{\phi}=\frac{1}{2}\omega(\phi)\dot{\phi}^2-V(\phi)\, ,
\end{equation}
and in addition, any extra matter perfect fluids with equation of state $P_m=w_m\rho_m$, contribute to the total energy density and pressure as follows,
\begin{equation}\label{completeenergypressure}
P=P_m+P_{\phi},\,\,\,\rho=\rho_m+\rho_{\phi}\, .
\end{equation}
In view of the above and in conjunction with Eq. (\ref{presenrg}), the LQC-corrected kinetic term of the non-canonical scalar field is equal to,
 \begin{equation}\label{omega}
 \omega(\phi)\dot{\phi}^2=-2\frac{\dot{H}}{\kappa^2(1-2\frac{\rho}{\rho_c})}-(\rho_m+P_m),
\end{equation}
while the scalar potential reads,
\begin{equation}\label{potential}
V(\phi)=\rho+\frac{\dot{H}}{\kappa^2(1-2\frac{\rho}{\rho_c})}-\frac{\rho_m}{2}+\frac{P_m}{2}\, .
\end{equation}
Notice that in both cases, when the limit $\rho_c\to \infty$ is taken, the correct classical expressions are obtained. In view of Eq. (\ref{correctdef}), the scalar potential can be written as follows,
\begin{equation}\label{potential1}
V(\phi)= \frac{3H^2}{\kappa^2-\frac{\kappa ^2 \rho_c- \sqrt{-12 H^2 \kappa ^2 \rho_c+\kappa ^4 \rho_c^2}}{2 \rho_c}}+\frac{\dot{H}}{\kappa^2(1-2\frac{\rho}{\rho_c})}-\frac{\rho_m}{2}+\frac{P_m}{2}\, .
\end{equation}
Now for the reconstruction technique in the scalar-tensor case, we seek for solutions to the cosmological equations of the following form \cite{scalarrecon0,scalarrecon1,scalarrecon2},
\begin{equation}\label{solutionsoftheform}
\phi=t,\,\,\,H(t)=f(t)\, ,
\end{equation}
and in effect, the non-canonical scalar field kinetic term and the corresponding scalar potential read,
 \begin{equation}\label{omega22}
 \omega(\phi)=-\frac{2\dot{H}}{\kappa^2-\frac{\kappa ^2 \rho_c- \sqrt{-12 H^2 \kappa ^2 \rho_c+\kappa ^4 \rho_c^2}}{ \rho_c}}-(1+w)\rho_0e^{-3(1+w_m)\int f(t)\mathrm{d}t},
\end{equation}
\begin{equation}\label{potential122}
V(\phi)= \frac{3H^2}{\kappa^2-\frac{\kappa ^2 \rho_c- \sqrt{-12 H^2 \kappa ^2 \rho_c+\kappa ^4 \rho_c^2}}{2 \rho_c}}+\frac{\dot{H}}{\kappa^2-\frac{\kappa ^2 \rho_c- \sqrt{-12 H^2 \kappa ^2 \rho_c+\kappa ^4 \rho_c^2}}{ \rho_c}}-\frac{(1-w)}{2}\rho_0e^{-3(1+w_m)\int f(t)\mathrm{d}t}\, ,
\end{equation}
where we made use of the equation of state $P_m=w_m\rho_m$ for the perfect fluid matter fields. The equations (\ref{omega22}) and (\ref{potential122}) constitute basically the LQC-corrected classical reconstruction method of Refs. \cite{scalarrecon0,scalarrecon1,scalarrecon2}, and in the sections that follow, we shall make use of these in order to realize various cosmological scenarios. Finally, let us note that if the classical limit of Eqs. (\ref{omega22}) and (\ref{potential122}) is taken, we obtain,
\begin{equation}\label{omega22classlim}
 \omega(\phi)=-\frac{2\dot{H}}{\kappa^2}-(1+w)\rho_0e^{-3(1+w_m)\int f(t)\mathrm{d}t},
\end{equation}
\begin{equation}\label{potential122classlim}
V(\phi)= \frac{3H^2}{\kappa^2}+\frac{\dot{H}}{\kappa^2}-\frac{(1-w)}{2}\rho_0e^{-3(1+w_m)\int f(t)\mathrm{d}t}\, ,
\end{equation}
which are exactly the classical reconstruction solutions of Refs. \cite{scalarrecon0,scalarrecon1,scalarrecon2}.

\section{Inflationary and Bounce Cosmologies with LQC Scalar Fields}

By using the reconstruction formalism of the previous section, in this section we shall realize various inflationary and bouncing cosmology cosmological scenarios, in the context of LQC-corrected scalar-tensor gravity. We aim to find the kinetic term and the scalar potential in each case, and also the corresponding EoS parameter $w_{eff}$. Also in each case we shall find approximate expressions for the EoS parameter in various cosmic time limits of interest, mainly focusing on the late-time era in all cases, and for the inflationary scenarios on the early times, while for the bouncing cosmologies we focus on cosmic times near the corresponding bouncing point. For simplicity, we assume that no perfect matter fluids are present, apart from the scalar field, so $\rho_m=p_m=0$.

\subsection{The Intermediate Inflation}

We start off our analysis with the intermediate inflation scenario \cite{Barrow:1990td}, in which case the Hubble rate and the scale factor are,
\begin{equation}\label{ex1}
a(t)=e^{A\,t^n},\,\,\, H(t)=A n t^{n-1}\, ,
\end{equation}
where $0<n<1$ and also with $A>0$ and of course $t>0$. In view of Eq. (\ref{omega22}) in conjunction with  (\ref{solutionsoftheform}), the kinetic term $\omega (t)$ for the intermediate inflation scenario is equal to,
\begin{equation}\label{omegat}
\omega (\phi)=\frac{2 A (1-n) n \rho_c t^{n-2}}{\sqrt{\kappa ^2 \rho_c \left(\kappa ^2 \rho_c-12 A^2 n^2 t^{2 n-2}\right)}}\, ,
\end{equation}
which as it can be seen, is always positive since $0<n<1$, and in effect in this case the scalar field is a non-phantom one. In addition, the scalar potential realizing the intermediate inflation scenario (\ref{ex1}) is,
\begin{equation}\label{potex1}
V(\phi)= \frac{A n t^n \left(\left(6 A n t^n+n-1\right) \sqrt{\kappa ^2 \rho_c \left(\kappa ^2 \rho_c-12 A^2 n^2 t^{2 n-2}\right)}+\kappa ^2 (n-1) \rho_c\right)}{\kappa ^2 \left(t^2 \left(\sqrt{\kappa ^2 \rho_c \left(\kappa ^2 \rho_c-12 A^2 n^2 t^{2 n-2}\right)}+\kappa ^2 \rho_c\right)-12 A^2 n^2 t^{2 n}\right)}
\, .
\end{equation}
The corresponding EoS of Eq. (\ref{eosdefcorrect}) can be easily calculated for the intermediate inflation case, and it reads,
\begin{equation}\label{bott}
w_{eff}=\frac{\kappa ^2 (1-n) \rho_c t^{-n}}{3 A n \sqrt{\kappa ^2 \rho_c \left(\kappa ^2 \rho_c-12 A^2 n^2 t^{2 n-2}\right)}}+\frac{(1-n) t^{-n}}{3 A n}-1\, .
\end{equation}
Let us investigate the behavior of the EoS for early-times, which correspond to the inflationary case for $t$ very small, and also for late times, which correspond to the present epoch. At early times, the EoS $w_{eff}$ (\ref{bott}) blows up for $t\to 0$. This is an expected behavior since the intermediate inflation scenario has a Type III singularity (see a later section for the classification of finite-time singularities) at the origin $t=0$. At late times that satisfy $\kappa ^2 \rho_c>12 e^{2 A t^n}$, the last two terms in (\ref{bott}) are subdominant and very small. Hence, at late times the EoS describe a nearly de Sitter evolution, slightly turned into the quintessential evolution, that is $w_{eff}\succeq -1$. This was expected, since in the LQC-corrected intermediate inflation scenario, the non-canonical scalar field is found to be a non-phantom one.

\subsection{The Symmetric Bounce Scenario}

Now let us turn our focus to the symmetric bounce scenario, in which case the scale factor and the Hubble rate are,
\begin{equation}\label{ex11}
a(t)=\exp \left(A t^2\right),\,\,\,H(t)=2 A t\, ,
\end{equation}
with  $A>0$. In effect, by combining Eqs. (\ref{omega22}) and (\ref{solutionsoftheform}), the kinetic term $\omega (t)$ becomes,
\begin{equation}\label{omegat1}
\omega (\phi)=-\frac{4 A \rho_c}{\sqrt{\kappa ^2 \rho_c \left(\kappa ^2 \rho_c-48 A^2 \phi^2\right)}}
\, ,
\end{equation}
hence in this case, the non-canonical scalar field is a phantom scalar field. Accordingly, the scalar potential reads,
\begin{equation}\label{potex11}
V(\phi)=\frac{2 A \left(\left(12 A \phi^2+1\right) \sqrt{\kappa ^2 \rho_c \left(\kappa ^2 \rho_c-48 A^2 \phi^2\right)}+\kappa ^2 \rho_c\right)}{\kappa ^2 \left(\sqrt{\kappa ^2 \rho_c \left(\kappa ^2 \rho_c-48 A^2 \phi^2\right)}-48 A^2 \phi^2+\kappa ^2 \rho_c\right)}
\, .
\end{equation}
Now let us turn our focus on the EoS parameter of Eq. (\ref{eosdefcorrect}), which in this case reads,
\begin{equation}\label{bott1}
w_{eff}=-\frac{\kappa ^2 \rho_c}{6 A t^2 \sqrt{\kappa ^2 \rho_c \left(\kappa ^2 \rho_c-48 A^2 t^2\right)}}-\frac{1}{6 A t^2}-1
\, ,
\end{equation}
and let us investigate its behavior for various cases of interest. As it can be seen, for cosmic times near the bouncing point $t=0$, the EoS becomes singular in this case, hence although the cosmological evolution of the symmetric bounce (\ref{ex11}) has no singularities, the EoS becomes singular at the origin. At late times, the EoS is approximately a nearly de Sitter evolution, slightly turned into phantom, like in the previous scenario of intermediate inflation. Hence this model may also describe a slightly turned into phantom de Sitter dark energy era.


\subsection{The Type IV Singular de Sitter Inflationary Scenario}

Now let us investigate the cosmological evolution for which a Type IV singularity occurs for a de Sitter evolution \cite{Odintsov:2015gba}. Let us note that this scenario is the most physically appealing of all the scenarios we shall study in this paper. This scenario was investigated in the context of $F(R)$ gravity in Ref. \cite{Odintsov:2015gba} and as it was demonstrated, the Universe may smoothly pass through the singularity, since the physical quantities defined on the three dimensional spacelike hypersurface are finite. The only effect that the singularity has is on the dynamics of inflation, see Ref. \cite{Odintsov:2015gba} for details. Also, similar possibilities occur in fluid cosmology, when soft singularities like the Type IV are used, see \cite{Brevik:2016kuy} for a characteristic study.

The cosmic evolution which describes a Type IV singular de Sitter evolution, has the following Hubble rate,
\begin{equation}\label{nbr1}
H(t)=H_0+f_0(t-t_s)^{\alpha}\, ,
\end{equation}
with $\alpha>1$ and $f_0>0$. The finite time singularity occurs at $t=t_s$ and depending on the values of $\alpha$, the following singularities may occur \cite{Nojiri:2005sx}:
\begin{itemize}\label{lista}
\item For $\alpha>1$, a Type IV singularity is realized at $t_s$.
\item For $0<\alpha<1$, a Type II singularity  is realized at $t_s$.
\item For $-1<\alpha<0$, a Type III singularity  is realized at $t_s$.
\item For $\alpha<-1$, a Type I, or so called Big Rip singularity  is realized at $t_s$,
\end{itemize}
We shall be interested in the case $\alpha>1$, and also we assume that $\alpha$ has the form $\alpha=2n/(2m+1)$, where $n$, $m$ are positive integers, in order to avoid complex values in the Hubble rate. The kinetic term of the LQC-corrected scalar-tensor theory that realizes the cosmological evolution (\ref{nbr1}) is,
\begin{equation}\label{omegacl}
\omega (\phi)= -\frac{2 \alpha  f_0 \rho_c (\phi-t_s)^{\alpha -1}}{\sqrt{\kappa ^2 \rho_c \left(\kappa ^2 \rho_c-12 \left(f_0 (\phi-t_s)^{\alpha }+H_0\right)^2\right)}}\, ,
\end{equation}
so for cosmic times prior to the singularity, the scalar field is a non-phantom one, while for cosmic times after the singularity, the scalar field becomes a phantom one. In principle, the singularity time instance may be chosen to occur at late times, for example it may occur at $t=t_s\simeq 13$Gy, so it is realized at the present epoch. Hence, in the context of this scenario, the scalar field is non-phantom until some time in the present epoch, and after the singularity, the scalar field becomes a phantom one. The corresponding scalar potential reads,
\begin{equation}\label{potcl}
V(\phi )=
\rho_c \left(\frac{\alpha  f_0 (\phi -t_s)^{\alpha -1}}{\sqrt{\kappa ^2 \rho_c \left(\kappa ^2 \rho_c-12 \left(f_0 (\phi -t_s)^{\alpha }+H_0\right)^2\right)}}+\frac{6 \left(f_0 (\phi -t_s)^{\alpha }+H_0\right)^2}{\sqrt{\kappa ^2 \rho_c \left(\kappa ^2 \rho_c-12 \left(f_0 (\phi -t_s)^{\alpha }+H_0\right)^2\right)}+\kappa ^2 \rho_c}\right)\, .
\end{equation}
Let us now investigate the behavior of the EoS which as we demonstrate has an interesting behavior. The EoS parameter $w_{eff}$ in this case is,
\begin{equation}\label{eoscl}
w_{eff}=-\frac{\alpha  f_0 \kappa ^2 \rho_c (t-t_s)^{\alpha -1}}{3 \left(f_0 (t-t_s)^{\alpha }+H_0\right)^2 \sqrt{\kappa ^2 \rho_c \left(\kappa ^2 \rho_c-12 \left(f_0 (t-t_s)^{\alpha }+H_0\right)^2\right)}}-\frac{\alpha  f_0 (t-t_s)^{\alpha -1}}{3 \left(f_0 (t-t_s)^{\alpha }+H_0\right)^2}-1
\, .
\end{equation}
For $t\ll t_s$, due to the values that the parameter $\alpha$ takes, the EoS becomes,
\begin{equation}\label{esos}
w_{eff}\simeq \frac{\alpha  \kappa ^2 \rho_c t_s^{\alpha -1}}{3 \left(H_0+t_s^{\alpha }\right)^2 \sqrt{\kappa ^2 \rho_c \left(\kappa ^2 \rho_c-12 \left(H_0+t_s^{\alpha }\right)^2\right)}}+\frac{\alpha  t_s^{\alpha -1}}{3 \left(H_0+t_s^{\alpha }\right)^2}-1/, ,
\end{equation}
which describes a nearly de Sitter evolution, slightly turned to the quintessential evolution. This result was expected since prior to the singularity the scalar field is non-phantom and for early times the evolution is de Sitter. Now, at late times near the singularity, the EoS is described by the following expression,
\begin{equation}\label{quasieosevolutionatlatetimes}
w_{eff}\simeq  -\frac{\alpha  \kappa ^2 \rho_c (t-t_s)^{\alpha -1}}{3 H_0^2 \sqrt{\kappa ^2 \rho_c \left(\kappa ^2 \rho_c-12 H_0^2\right)}}-\frac{\alpha  (t-t_s)^{\alpha -1}}{3 H_0^2}-1\, .
\end{equation}
The above EoS before the singularity describes a de Sitter late-time era, slightly turned to a quintessential phase, while after the singularity it describes a de Sitter late-time era, slightly turned to a phantom evolution. This is particularly interesting, since in this scenario, the early and late-time acceleration era may be described in a unified way by the same model. More interestingly, during the dark energy era, a transition from non-phantom to a phantom dark energy era occurs.

\subsection{The Superbounce Scenario}

The last scenario we shall realize in the context of LQC-corrected scalar field theory is the superbounce scenario \cite{Odintsov:2015uca}, in which case, the scale factor and the corresponding Hubble rate are,
\begin{equation}\label{superbouncescalehubblerate}
a(t)=(t_s-t)^{\frac{2}{c^2}},\,\,\,H(t)=\frac{2}{c^2 (t-t_s)}\, ,
\end{equation}
where $t=t_s$ is the time instance that the bouncing point occurs, while $c<\sqrt{6}$. In this case, the kinetic term of the non-canonical scalar field is,
\begin{equation}\label{omegaclsuperbounce}
\omega (\phi)= \frac{4 \rho_c}{c^2 (\phi-t_s)^2 \sqrt{\kappa ^2 \rho_c \left(\kappa ^2 \rho_c-\frac{48}{c^4 (\phi-t_s)^2}\right)}}\, ,
\end{equation}
and as it can be seen, prior to the bouncing point, the kinetic term is positive and for $t>t_s$, so the scalar field is a non-phantom one for all cosmic times. Also in this case, the theory has a structural instability at the Lagrangian level, exactly at the bouncing point, since the kinetic term becomes singular at $t=t_s$. Accordingly, the scalar potential is in this case,
\begin{equation}\label{scalarpotentialsuperbounce}
V(\phi)=\frac{2 \rho_c \left(\frac{12}{\sqrt{\kappa ^2 \rho_c \left(\kappa ^2 \rho_c-\frac{48}{c^4 (t_s-\phi )^2}\right)}+\kappa ^2 \rho_c}-\frac{c^2}{\sqrt{\kappa ^2 \rho_c \left(\kappa ^2 \rho_c-\frac{48}{c^4 (t_s-\phi )^2}\right)}}\right)}{c^4 (t_s-\phi )^2}\, .
\end{equation}
Finally, the LQC-corrected EoS parameter $w_{eff}$ in the superbounce case reads,
\begin{equation}\label{eossuperbouncelqc}
w_{eff}=\frac{c^4 \kappa ^2 \rho_c \left| t-t_s\right| }{6 \sqrt{\kappa ^2 \rho_c \left(c^4 \kappa ^2 \rho_c (t-t_s)^2-48\right)}}+\frac{c^2}{6}-1\, ,
\end{equation}
which is non-singular at $t=t_s$. It is worth investigating the behavior of the EoS for various cosmic times. First of all, the EoS parameter is quintessential for all cosmic times, and this result is independent from whether $t>t_s$ or $t<t_s$. This behavior was expected, since from Eq. (\ref{omegaclsuperbounce}) we concluded that for all cosmic times, the scalar field is non-phantom. Secondly, at late times, the EoS  parameter becomes approximately,
\begin{equation}\label{eosapprox}
w_{eff}\simeq \frac{c^2}{3}-1\, ,
\end{equation}
so the evolution is accelerating for all the allowed values of $c<\sqrt{6}$, as long as $w_{eff}<-\frac{1}{3}$. Therefore, for this scenario too, the late-time era has an accelerating EoS, so the dark energy era may also be described in this case.

What now remains to conclude our study, is to check the dynamical stability of the solutions (\ref{solutionsoftheform}) at the level of cosmological equations. What we will actually do is to perturb the cosmological dynamical system for all the solutions we found, and we investigate the stability of the dynamical system, by keeping linear perturbation terms. This is the subject of the next section.


\section{Dynamical Stability of the Solutions at First Order}

In the previous section we investigated how it is possible to realize various cosmological evolutions in the context of LQC-corrected scalar-tensor gravity, by using solutions of the form (\ref{solutionsoftheform}). However, the solution (\ref{solutionsoftheform}) is not the only solution of the cosmological dynamical system under study, so it is compelling to investigate whether this system is stable or not. In view of this issue, in this section we shall investigate whether the solution (\ref{solutionsoftheform}) is stable towards linear perturbations of the cosmological equations. The cosmological equations of the system at hand are Eqs. (\ref{eqnm}) and (\ref{classicaleqns}), and in addition, there is another equation for the non-canonical scalar field, which is,
\begin{equation}\label{firledobey}
\omega (\phi)\ddot{\phi}+\frac{1}{2}\omega '(\phi)\dot{\phi}^2+3H\omega (\phi)\dot{\phi}+V'(\phi)=0\, ,
\end{equation}
with the prime denoting differentiation with respect to the non-canonical scalar field $\phi$. In order to render the study of the cosmological dynamical system easier, we shall use the following variables,
\begin{equation}\label{dynsvars}
X=\dot{\phi},\,\,\,Y=\frac{f(\phi )}{H}\, ,
\end{equation}
and in effect, the cosmological equations (\ref{eqnm}), (\ref{classicaleqns}) and (\ref{dynsvars}) can be expressed in terms of the new variables (\ref{dynsvars}),
\begin{align}\label{perteqns}
& \frac{\mathrm{d}X}{\mathrm{d}N}=-3 (X-Y)+\frac{6 Y f'(t)}{\kappa ^2 \rho_c}+\frac{Y \left(9 f(t)^2-6 X^2 f'(t)\right)}{\kappa ^2 \rho_c}+\frac{f''(t)}{H(t) f'(t)}-\frac{X^2 f''(t)}{2 H(t) f'(t)}\\ \notag &
\frac{\mathrm{d}Y}{\mathrm{d}N}=\frac{X f'(t)}{H(t)^2}-\frac{X^2 Y f'(t)}{H(t)^2}\, .
\end{align}
Note that for the dynamical system (\ref{perteqns}) we expressed the variables as functions of the $e$-foldings number $N=\ln a$, and also we used the differentiation rule $\frac{\mathrm{d}}{\mathrm{d}N}=H^{-1}\frac{\mathrm{d}}{\mathrm{d}t}$. In the classical limit $\rho_c\to \infty$, the dynamical system takes its classical form \cite{scalarrecon3},
\begin{align}\label{perteqnsperitto}
& \frac{\mathrm{d}X}{\mathrm{d}N}=-3 (X-Y)+\frac{f''(t)}{H(t) f'(t)}-\frac{X^2 f''(t)}{2 H(t) f'(t)}\\ \notag &
\frac{\mathrm{d}Y}{\mathrm{d}N}=\frac{X f'(t)}{H(t)^2}-\frac{X^2 Y f'(t)}{H(t)^2}\, .
\end{align}

The solution (\ref{solutionsoftheform}), in terms of the variables $X$ and $Y$, takes the values $(X,Y)=(1,1)$, so the stability of the solution $(X,Y)=(1,1)$, can be examined if the following linear perturbations are performed,
\begin{equation}\label{pertactual}
X=1+\delta X,\,\,\,Y=1+\delta Y\, .
\end{equation}
By combining  (\ref{pertactual}) and (\ref{perteqns}), at linear order of the variations $\delta X$ and $\delta Y$, we obtain the following dynamical evolution,
\begin{equation}\label{perteqn2}
\left(
\begin{array}{c}
 \frac{\mathrm{d} X}{\mathrm{d}N} \\
 \frac{\mathrm{d} Y}{\mathrm{d} N} \\
\end{array}
\right)=\left(
\begin{array}{cc}
 -3-\frac{ H''(t)}{H(t) H'(t)}-\frac{12 H'(t)}{\kappa ^2 \rho_c} & 3+\frac{9 H(t)^2}{\kappa ^2 \rho_c} \\
 -\frac{H'(t)}{H(t)^2} & -\frac{H'(t)}{H(t)^2} \\
\end{array}
\right)\left(
\begin{array}{c}
 \delta X \\
 \delta Y \\
\end{array}
\right)\, .
\end{equation}
The above evolution will determine the evolution of the perturbations of the solution $(X,Y)=(1,1)$. Note that if the classical limit is taken, the above dynamical evolution becomes,
\begin{equation}\label{perteqn2peritto}
\left(
\begin{array}{c}
 \frac{\mathrm{d} X}{\mathrm{d}N} \\
 \frac{\mathrm{d} Y}{\mathrm{d} N} \\
\end{array}
\right)=\left(
\begin{array}{cc}
 -3-\frac{ H''(t)}{H(t) H'(t)} & 3 \\
 -\frac{H'(t)}{H(t)^2} & -\frac{H'(t)}{H(t)^2} \\
\end{array}
\right)\left(
\begin{array}{c}
 \delta X \\
 \delta Y \\
\end{array}
\right)\, ,
\end{equation}
which is identical to the dynamical system of Ref. \cite{scalarrecon3}. Returning to the study of the dynamical system (\ref{perteqn2}), the stability can be determined once we know the eigenvalues $m_{1,2}$ of the matrix $M$, which appears in Eq. (\ref{perteqn2}), namely,
\begin{equation}\label{m}
M=\left(
\begin{array}{cc}
 -3-\frac{ H''(t)}{H(t) H'(t)}-\frac{12 H'(t)}{\kappa ^2 \rho_c} & 3+\frac{9 H(t)^2}{\kappa ^2 \rho_c} \\
 -\frac{H'(t)}{H(t)^2} & -\frac{H'(t)}{H(t)^2} \\
\end{array}
\right)\, ,
\end{equation}
with the explicit form of the eigenvalues is given below,
\begin{align}\label{m1}
& m_{1,2}=\frac{1}{2 \kappa ^2 \rho_c H(t)^3 H'(t)}\left(-3 \kappa ^2 \rho_c H(t)^3 H'(t)-\kappa ^2 \rho_c H(t) H'(t)^2-12 H(t)^3 H'(t)^2-\kappa ^2 \rho_c H(t)^2 H''(t)\pm \sqrt{Q(t)}\right)
\end{align}
where $Q(t)$ is,
\begin{align}\label{qt}
& Q(t)=\left(3 \kappa ^2 \rho_c H(t)^3 H'(t)+\kappa ^2 \rho_c H(t) H'(t)^2+12 H(t)^3 H'(t)^2+\kappa ^2 \rho_c H(t)^2 H''(t)\right)^2\\ \notag & -4 \kappa ^2 \rho_c H(t)^3 H'(t) \left(6 \kappa ^2 \rho_c H(t) H'(t)^2+9 H(t)^3 H'(t)^2+12 H(t) H'(t)^3+\kappa ^2 \rho_c H'(t) H''(t)\right)\, .
\end{align}
About the stability of the system, when the eigenvalues are positive, the solution  (\ref{solutionsoftheform}) is stable against linear perturbations. If one of the two eigenvalues is positive, then the dynamical system is unstable, and if one of the two eigenvalues is zero, then no conclusion can be made for the stability of the solution (\ref{solutionsoftheform}). Our task in this section is to investigate the stability of the dynamical equations at linear order, by examining the eigenvalues (\ref{m1}) for each cosmological evolution, and we compare the LQC-corrected case with the classical case.
Let us start our analysis with the intermediate inflation scenario (\ref{ex1}), in which case the matrix appearing in Eq. (\ref{perteqn2}) is equal to,
\begin{equation}\label{matrixintermediate}
M=\left(
\begin{array}{cc}
 -\frac{12 A (n-1) n t^{n-2}}{\kappa ^2 \rho_c}-\frac{(n-2) t^{-n}}{A n}-3 & \frac{9 A^2 n^2 t^{2 n-2}}{\kappa ^2 \rho_c}+3 \\
 -\frac{(n-1) t^{-n}}{A n} & -\frac{(n-1) t^{-n}}{A n} \\
\end{array}
\right)\, ,
\end{equation}
the eigenvalues of which are,
\begin{align}\label{eigeninter}
& m_1= \frac{1}{2 A^4 \kappa ^2 (n-1) n^4 \rho_c}t^{-4 n-2}\left(-12 A^5 (n-1)^2 n^5 t^{5 n}-A^3 \kappa ^2 n^3 \left(2 n^2-5 n+3\right) \rho_c t^{3 n+2}\right. \\ \notag &
\frac{1}{2 A^4 \kappa ^2 (n-1) n^4 \rho_c}t^{-4 n-2}\left(-12 A^5 (n-1)^2 n^5 t^{5 n}-A^3 \kappa ^2 n^3 \left(2 n^2-5 n+3\right) \rho_c t^{3 n+2}\right.\\ \notag &
+36 A^3 \kappa ^2 (n-1) n^3 \rho_c t^{3 n+2}+\kappa ^4 \rho_c^2 t^4 \\ \notag &
\left.\left.\left.-3 A^2 \kappa ^2 n^2 \rho_c t^{2 n+2} \left(8 n-3 \kappa ^2 \rho_c t^2-8\right)-6 A \kappa ^4 n (2 n-1) \rho_c^2 t^{n+4}\right)\right)\right)\\ \notag &
m_2=\frac{1}{2 A^4 \kappa ^2 (n-1) n^4 \rho_c}t^{-4 n-2}\left(-12 A^5 (n-1)^2 n^5 t^{5 n}-A^3 \kappa ^2 n^3 \left(2 n^2-5 n+3\right) \rho_c t^{3 n+2}\right. \\ \notag &
\frac{1}{2 A^4 \kappa ^2 (n-1) n^4 \rho_c}t^{-4 n-2}\left(-12 A^5 (n-1)^2 n^5 t^{5 n}-A^3 \kappa ^2 n^3 \left(2 n^2-5 n+3\right) \rho_c t^{3 n+2}\right.\\ \notag &
\left.\left.\left.-3 A^2 \kappa ^2 n^2 \rho_c t^{2 n+2} \left(8 n-3 \kappa ^2 \rho_c t^2-8\right)-6 A \kappa ^4 n (2 n-1) \rho_c^2 t^{n+4}+\kappa ^4 \rho_c^2 +t^4\right)\right)\right)\, .
\end{align}
In order to see whether the eigenvalues are positive, zero or negative, we need to specify the values of the parameters, since in general it is not easy to have a concrete idea of how the eigenvalues behave. So we assume that the free parameters take the following values,
\begin{equation}\label{prmetersvaluesgeneralintermediate}
\kappa^2\rho_c=10^{12},\,\,\,n=0.9,\,\,\,A=1,
\end{equation}
so at early times, for example if $t\simeq 10^{-35}$sec, the eigenvalues read,
\begin{equation}\label{eogevaluesaeearlyintermediate}
m_1\simeq 5.76\times 10^{17},\,\,\,m_2\simeq 10^{34}\, ,
\end{equation}
while at late times these become,
\begin{equation}\label{eogevaluesaeearlyintermediatelate}
m_1\simeq -3.898\times 10^{1715},\,\,\,m_2\simeq 1.773 \times 10^{1703}\, .
\end{equation}
Therefore, in both cases the solution is unstable (\ref{solutionsoftheform}), which means that the dynamical system will reach this solution but it will abandon it eventually at finite time. Also, the classical result of this scenario also predicts instability of the solution (\ref{solutionsoftheform}), as it can be checked in a similar manner.

Let us now proceed to the symmetric bounce case of Eq. (\ref{ex11}), in which case the matrix $M$ reads,
\begin{equation}\label{matrixintermediate1}
M=\left(
\begin{array}{cc}
 -\frac{24 A}{\kappa ^2 \rho_c}-3 & \frac{36 A^2 t^2}{\kappa ^2 \rho_c}+3 \\
 -\frac{1}{2 A t^2} & -\frac{1}{2 A t^2} \\
\end{array}
\right)\, ,
\end{equation}
the eigenvalues of which are,
\begin{align}\label{eigeninter1}
& m_{1,2}=\frac{1}{4} \left(\mp\frac{\sqrt{A^6 t^2 \left(2304 A^4 t^4+288 A^3 \kappa ^2 \rho_c t^4+12 A^2 \kappa ^2 \rho_c t^2 \left(3 \kappa ^2 \rho_c t^2-8\right)-36 A \kappa ^4 \rho_c^2 t^2+\kappa ^4 \rho_c^2\right)}}{A^4 \kappa ^2 \rho_c t^3}-\frac{48 A}{\kappa ^2 \rho_c}-\frac{1}{A t^2}-6\right)
\, .
\end{align}
As in the previous case, we specify the values of the free parameters as follows,
\begin{equation}\label{prmetersvaluesgeneralintermediate1}
\kappa^2\rho_c=10^{12},\,\,\,A=1,
\end{equation}
so as in the previous case if $t\simeq 10^{-35}$sec, the eigenvalues read,
\begin{equation}\label{eogevaluesaeearlyintermediate1}
m_1\simeq -\times 10^{35},\,\,\,m_2\simeq -9.22\times 10^{18}\, ,
\end{equation}
while at late times the eigenvalues are,
\begin{equation}\label{eogevaluesaeearlyintermediatelate1}
m_1\simeq -7.279\times 10^{740},\,\,\,m_2\simeq -1.5\, .
\end{equation}
Hence for the symmetric bounce scenario, the cosmological equations are stable towards linear perturbations, therefore the solution (\ref{solutionsoftheform}) is stable. It is notable that in the classical case, the eigenvalues are negative at early times, so the system is stable classically, however at late times the eigenvalue $m_2$ is zero, which means that the LQC-corrected symmetric bounce is stable even at late times, in contrast to the classical picture.

Let us now turn our focus on the Type IV singular de Sitter evolution, in which case, the matrix $M$ is,
\begin{equation}\label{matrixintermediate1typiv}
M=\left(
\begin{array}{cc}
 -\frac{12 f_0 \alpha  (t-t_s)^{\alpha -1}}{\kappa ^2 \rho_c}-\frac{\alpha -1}{\left(f_0 (t-t_s)^{\alpha }+H_0\right) (t-t_s)}-3 & \frac{9 \left(f_0 (t-t_s)^{\alpha }+H_0\right)^2}{\kappa ^2 \rho_c}+3 \\
 -\frac{f_0 (t-t_s)^{\alpha -1} \alpha }{\left(f_0 (t-t_s)^{\alpha }+H_0\right)^2} & -\frac{f_0 (t-t_s)^{\alpha -1} \alpha }{\left(f_0 (t-t_s)^{\alpha }+H_0\right)^2} \\
\end{array}
\right)\, ,
\end{equation}
the eigenvalues of which are,
\begin{align}\label{eigeninter1typiv}
& m_{1}=\frac{1}{2}\left(-\frac{\alpha  f_0 (t-t_s)^{\alpha -1}}{\left(f_0 (t-t_s)^{\alpha }+H_0\right)^2}-\frac{\alpha -1}{(t-t_s) \left(f_0 (t-t_s)^{\alpha }+H_0\right)}-3\right.\\ \notag & -\frac{(t-t_s)^{1-\alpha }}{\alpha  f_0 \kappa ^2 \rho_c \left(f_0 (t-t_s)^{\alpha }+H_0\right)^3}-\frac{12 \alpha  f_0 (t-t_s)^{\alpha -1}}{\kappa ^2 \rho_c}\\ \notag &
\surd \left(\alpha ^2 f_0^2 (t-t_s)^{2 \alpha -4} \left(f_0 (t-t_s)^{\alpha }+H_0\right)^2-4 \alpha  f_0 \kappa ^2 \rho_c (t-t_s)^{\alpha } \left(f_0 (t-t_s)^{\alpha }+H_0\right)\right.\\ \notag &
\left(12 \alpha  f_0 (t-t_s)^{\alpha } \left(f_0 (t-t_s)^{\alpha }+H_0\right)+9 (t-t_s) \left(f_0 (t-t_s)^{\alpha }+H_0\right)^3\right.\\ \notag &
\left.(\alpha -1) \kappa ^2 \rho_c++6 \kappa ^2 \rho_c (t-t_s) \left(f_0 (t-t_s)^{\alpha }+H_0\right)\right)\\ \notag & +\left(3 \kappa ^2 \rho_c (t-t_s) \left(f_0 (t-t_s)^{\alpha }+H_0\right)^2+12 \alpha  f_0 (t-t_s)^{\alpha } \left(f_0 (t-t_s)^{\alpha }+H_0\right)^2\right. \\ \notag & \left.\left.\left.\left.(\alpha -1) \kappa ^2 \rho_c +\left(f_0 (t-t_s)^{\alpha }+H_0\right)+\alpha  f_0 \kappa ^2 \rho_c (t-t_s)^{\alpha }\right)^2\right)\right)\right)\\ \notag &
m_2= \left.\left.\left.\left.(\alpha -1) \kappa ^2 \rho_c +\left(f_0 (t-t_s)^{\alpha }+H_0\right)+\alpha  f_0 \kappa ^2 \rho_c (t-t_s)^{\alpha }\right)^2\right)\right)\right)\\ \notag & \frac{1}{2}\left(-\frac{\alpha  (t-t_s)^{\alpha -1}}{\left(H_0+(t-t_s)^{\alpha }\right)^2}-\frac{\alpha -1}{(t-t_s) \left(H_0+(t-t_s)^{\alpha }\right)}-\frac{12 \alpha  (t-t_s)^{\alpha -1}}{\kappa ^2 \rho_c}-3\right.\\ \notag & \frac{1}{2}\left(-\frac{\alpha  (t-t_s)^{\alpha -1}}{\left(H_0+(t-t_s)^{\alpha }\right)^2}-\frac{\alpha -1}{(t-t_s) \left(H_0+(t-t_s)^{\alpha }\right)}-\frac{12 \alpha  (t-t_s)^{\alpha -1}}{\kappa ^2 \rho_c}-3\right. \\ \notag & +\left.(\alpha -1) \kappa ^2 \rho_c+6 \kappa ^2 \rho_c (t-t_s) \left(H_0+(t-t_s)^{\alpha }\right)\right)\\ \notag  & \left(3 \kappa ^2 \rho_c (t-t_s) \left(H_0+(t-t_s)^{\alpha }\right)^2+12 \alpha  (t-t_s)^{\alpha } \left(H_0+(t-t_s)^{\alpha }\right)^2\right. \\ \notag & \left.\left.\left.\left.(\alpha -1) \kappa ^2 \rho_c +\left(H_0+(t-t_s)^{\alpha }\right)+\alpha  \kappa ^2 \rho_c (t-t_s)^{\alpha }\right)^2\right)\right)\right)
\, .
\end{align}
Let us now see the behavior of the eigenvalues at early times, so assume that the free parameters are chosen as follows,
\begin{equation}\label{prmetersvaluesgeneralintermediate1typiv}
\kappa^2\rho_c=10^{12},\,\,\,H_0=10^{13}\mathrm{sec}^{-1},\,\,\,\alpha=2.3,\,\,\,t_s=10^{17},\,\,\,f_0=10^{-8}\, ,
\end{equation}
so as in the previous case if we choose $t\simeq 10^{-35}$sec, the eigenvalues become equal to,
\begin{equation}\label{eogevaluesaeearlyintermediate1typiv}
m_1\simeq -2.76\times 10^{13},\,\,\,m_2\simeq -4.1943\times 10^{6}\, .
\end{equation}
As it can be checked, the classical case is unstable, hence the LQC corrections stabilize the cosmological solution of the Type IV singular de Sitter Universe. Finally let us discuss the superbounce case, for which the matrix $M$ reads,
\begin{equation}\label{matrixintermediate1typiv}
M=\left(
\begin{array}{cc}
 c^2+\frac{24}{c^2 (t-t_s)^2 \kappa ^2 \rho_c}-3 & 3+\frac{36}{c^4 (t-t_s)^2 \kappa ^2 \rho_c} \\
 \frac{c^2}{2} & \frac{c^2}{2} \\
\end{array}
\right)\, ,
\end{equation}
and the corresponding eigenvalues are,
\begin{align}\label{eigeninter1typiv}
& m_{1}=\frac{1}{4}\left(\frac{48}{c^2 \kappa ^2 \rho_c (t-t_s)^2}+3 c^2-6\right.\\ \notag &
\left.+\frac{c^8 (t-t_s)^5 \sqrt{\frac{c^8 \kappa ^4 \rho_c^2 (t-t_s)^4+12 c^6 \kappa ^4 \rho_c^2 (t-t_s)^4+36 c^4 \kappa ^4 \rho_c^2 (t-t_s)^4+96 c^4 \kappa ^2 \rho_c (t-t_s)^2-288 c^2 \kappa ^2 \rho_c (t-t_s)^2+2304}{c^{20} (t-t_s)^{14}}}}{\kappa ^2 \rho_c}\right),
\\ \notag &
m_2= \frac{12}{c^2 \kappa ^2 \rho_c (t-t_s)^2}+\frac{3 c^2}{4}-\frac{3}{2}\\ \notag &
-\frac{c^8 (t-t_s)^5 \sqrt{\left(\frac{384}{c^{10} (t-t_s)^7}-\frac{48 \kappa ^2 \rho_c}{c^8 (t-t_s)^5}+\frac{24 \kappa ^2 \rho_c}{c^6 (t-t_s)^5}\right)^2+\frac{64 \kappa ^2 \rho_c \left(\frac{288}{c^{10} (t-t_s)^7}-\frac{192}{c^8 (t-t_s)^7}+\frac{48 \kappa ^2 \rho_c}{c^6 (t-t_s)^5}-\frac{8 \kappa ^2 \rho_c}{c^4 (t-t_s)^5}\right)}{c^8 (t-t_s)^5}}}{32 \kappa ^2 \rho_c}\, .
\end{align}
By choosing the free parameters as follows,
\begin{equation}\label{prmetersvaluesgeneralintermediate1typiv}
\kappa^2\rho_c=10^{12},\,\,\,c=1/\sqrt{2},\,\,\,t_s=1\mathrm{sec}\, ,
\end{equation}
the eigenvalues near the bouncing point $t=t_s$ become equal to,
\begin{equation}\label{eogevaluesaeearlyintermediate1typiv}
m_1\simeq -2.92845,\,\,\,m_2\simeq 1.3\times 10^{16}\, .
\end{equation}
In effect, the superbounce solution is unstable near the bouncing point, and also as it can be verified, in the classical approach, one has one positive and one negative eigenvalue, hence the classical system is also unstable.

In conclusion, the addition of LQC holonomy corrections in the scalar-tensor theories may in some cases bring stability to the cosmological solutions we described in the previous section. However, there is no obvious pattern on how this stability feature is possible and it seems that this behavior is model dependent.

\section{Conclusions}

In this work we used the LQC-corrected scalar-tensor reconstruction formalism in order to realize various cosmological evolutions and also to study the stability of the solutions. Particularly we realized two singular inflationary cosmologies, the intermediate inflation scenario and the Type IV singular de Sitter scenario, and also two bouncing cosmologies, namely the superbounce and the symmetric bounce. We provided the exact functional forms of the kinetic term and of the scalar potential of the corresponding LQC-corrected scalar theory which realizes the above cosmologies, and we investigated the behavior of the EoS, focusing on time instances and epochs of interest, depending on the model. Moreover, we investigated the stability of the solutions we found, and as we demonstrated, in some cases, the evolution might be rendered stable, a feature absent in the classical counterpart theory.

The study we performed in this paper can be extended in the case of multiple scalar fields, in which case the realization of various cosmologies can be done in an easier way. However, the use of two scalars might impose the condition that one of these is phantom, so this possibility should be carefully studied. We hope to address the above approach in a future work.

\section*{Acknowledgments}

Financial support by the Research
Committee of the Technological Education Institute of Central Macedonia, Serres,
under grant SAT/ME/011117-192/12, is gratefully acknowledged.

\end{document}